%% file: main.tex
\documentclass[journal]{IEEEtran}

\usepackage{cite}
\usepackage{amsmath,amssymb,amsfonts}
\usepackage{graphicx}
\usepackage{textcomp}
\usepackage{xcolor}
\usepackage{tikz} 
\usepackage{pgfplots} 
\usepackage{float}
\usepackage{longtable}
\usepackage{multirow}
\usepackage[hidelinks]{hyperref}
\usepackage{algorithm}
\usepackage{algpseudocode}
\usepackage{booktabs}
\usepackage{todonotes}
\usepackage{balance}
\usepackage{soul}		
\usepackage{dblfloatfix}
\usepackage{subcaption}
\usepackage{nameref}
\usepackage{todonotes}

\def\BibTeX{{\rm B\kern-.05em{\sc i\kern-.025em b}\kern-.08em
    T\kern-.1667em\lower.7ex\hbox{E}\kern-.125emX}}
    
\graphicspath{{./img/}}

\usepgfplotslibrary{units}

\pgfplotsset{compat = 1.3}

\algblock{ParFor}{EndParFor}
\algnewcommand\algorithmicparfor{\textbf{parfor}}
\algnewcommand\algorithmicpardo{\textbf{do}}
\algnewcommand\algorithmicendparfor{\textbf{end\ parfor}}
\algrenewtext{ParFor}[1]{\algorithmicparfor\ #1\ \algorithmicpardo}
\algrenewtext{EndParFor}{\algorithmicendparfor}

\newcommand{\acronimo}{DSB}

\newcommand{\papertitle}{HAPM - Hardware Aware Pruning Method for CNN hardware accelerators in resource constrained devices}

\usepackage{environ}
\NewEnviron{scaled_IEEEeqnarray}{%
	\begin{IEEEeqnarray}{l}
		\scalebox{0.8}{$\BODY$}
	\end{IEEEeqnarray}
}

\begin{document}



\title{\papertitle}


\author{
	\IEEEauthorblockN{Federico Nicolás Peccia\IEEEauthorrefmark{1},
		Luciano Ferreyro\IEEEauthorrefmark{2}, 
		Alejandro Furfaro\IEEEauthorrefmark{3}} \\
	\IEEEauthorblockA{Digital Processing Laboratory, Electronics Department\\
		Universidad Tecnológica Nacional\\
		Email: \IEEEauthorrefmark{1}peccfederico@frba.utn.edu.ar,
		\IEEEauthorrefmark{2}lferreyro@frba.utn.edu.ar,
		\IEEEauthorrefmark{3}afurfaro@frba.utn.edu.ar}}

	
\maketitle

\input{paper_body_v3.tex}

\balance

\bibliographystyle{IEEEtran}
\bibliography{bib}

\end{document}

%% file: paper_body_v3.tex
\begin{abstract}
During the last years, algorithms known as Convolutional Neural Networks (CNNs) had become increasingly popular, expanding its application range to several areas. In particular, the image processing field has experienced a remarkable advance thanks to this algorithms. In IoT, a wide research field aims to develop hardware capable of execute them at the lowest possible energy cost, but keeping acceptable image inference time. One can get around this apparently conflicting objectives by applying design and training techniques. The present work proposes a generic hardware architecture ready to be implemented on FPGA devices, supporting a wide range of configurations which allows the system to run different neural network architectures, dynamically exploiting the \textit{sparsity} caused by pruning techniques in the mathematical operations present in this kind of algorithms. The inference speed of the design is evaluated over different resource constrained FPGA devices. Finally, the standard pruning algorithm is compared against a custom pruning technique specifically designed to exploit the scheduling properties of this hardware accelerator. We demonstrate that our \textit{hardware-aware} pruning algorithm achieves a remarkable improvement of a 45 \% in inference time compared to a network pruned using the standard algorithm.
\end{abstract}
	

\section{Introduction}
\IEEEPARstart{A}{rtificial} neural networks can be defined as a succession of $n$ layers of \textit{neurons} interconnected in a sequential manner, where the output is given by:

\begin{IEEEeqnarray}{l}
		\theta = \left(\sum_{k=0}^{j} w_{k}*i_{k} \right) + bias \\
		s = f_{(\theta)}
\end{IEEEeqnarray}

Where $\theta$ is the weighted sum of the input of the layer plus a $bias$ value, $s$ refers to the output of the layer and $f$ is an activation function.

In a convolutional layer the output is the result of a convolution process between the inputs and a \textit{matrix of weights}. As each layer dispose of this matrices and a convolution process is performed, it's possible to think each layer as a filter and the weights as the filter's \textit{coefficients} (a \textit{kernel}). The increasingly complex networks that are still being developed based on this layers sometimes need to do millions or even billions of multiply and accumulate operations in order to process one single image.


For this reason, once trained, these algorithms can undergo different kinds of optimizations according to the hardware on which they will be executed. One of the most commonly used techniques is the absolute value \textit{pruning} \cite{DBLP:journals/corr/HanPTD15,DBLP:journals/corr/NarangDSE17,DBLP:journals/corr/SeeLM16}, which eliminates redundant weights. Therefore, at running an algorithm's forward pass, the number of calculations and memory accesses needed decreases significantly.

An evolution of this technique was proposed by Zhu M. et. al. \cite{zhu2017prune}, who proposed to gradually force coefficients of each layer to zero until the amount of zero coefficients reaches a certain \textit{sparsity} threshold while retraining to accommodate the remaining parameters and thus maintaining the quality of the prediction.


On the other hand, in hardware accelerator designs specifically developed to run a neural network in inference mode, the hardware resource and energy consumption savings are one of the main concerns.
This kind of development presents very different challenges from those found during neural network training since the main focus is to find the most efficient way of executing the internal operations needed for an inference pass.

\begin{figure*}
	\centering
	\begin{subfigure}{0.32\textwidth}
		\centering
		\includegraphics[width=\textwidth]{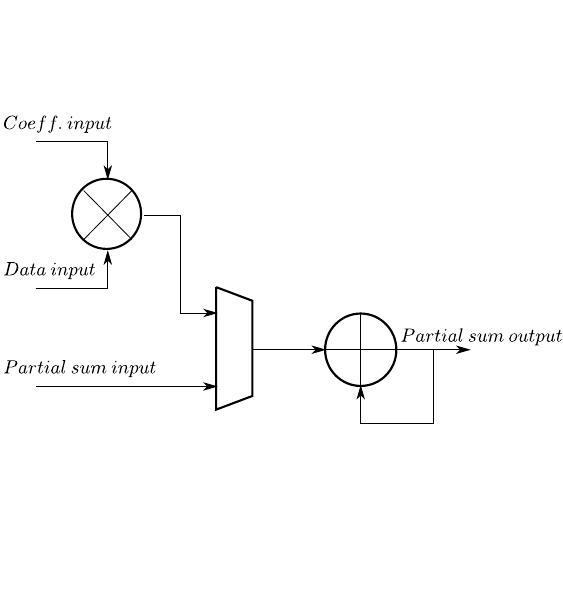}
		\caption{}
		\label{fig:sys_array_example_a}
	\end{subfigure}
	\hfill
	\begin{subfigure}{0.32\textwidth}
		\centering
		\includegraphics[width=\textwidth]{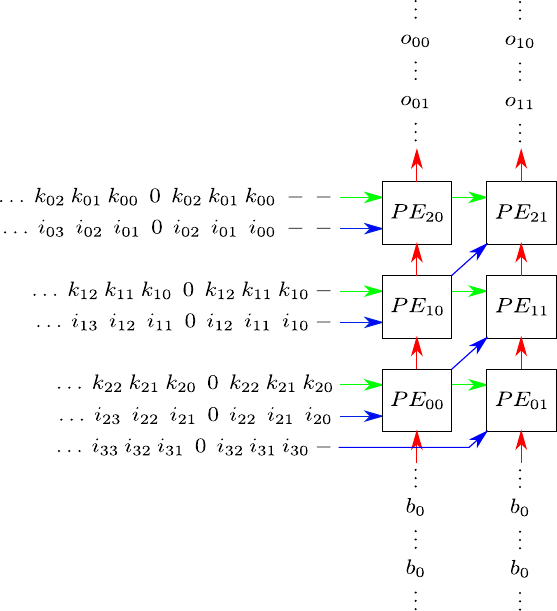}
		\caption{}
		\label{fig:sys_array_example_b}
	\end{subfigure}
	\hfill
	\begin{subfigure}{0.32\textwidth}
		\centering
		\includegraphics[width=\textwidth]{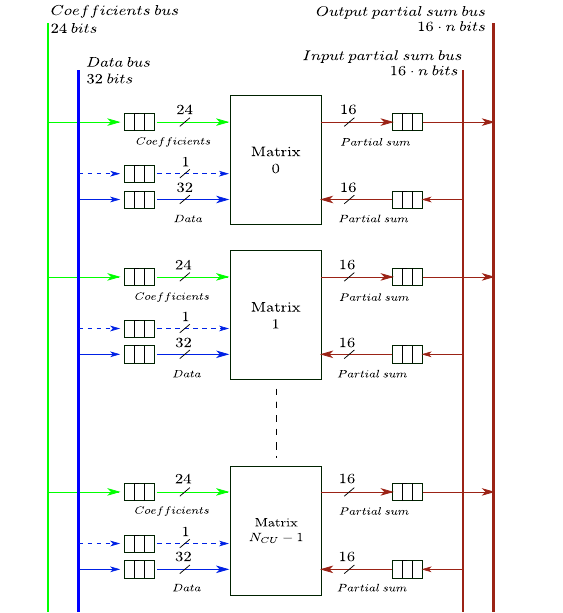}
		\caption{}
		\label{fig:sys_array_example_c}
	\end{subfigure}
	\caption{Simplified block diagram of a PE used in the design (\protect\subref{fig:sys_array_example_a}), scheduling of a single $3\times3$  kernel convolution on a computation units matrix of $CU_x = 2$ and $CU_y = 3$ (\protect\subref{fig:sys_array_example_b}) and $N_{CU}$ computation units matrices sharing data, kernel and partial sum buses (\protect\subref{fig:sys_array_example_c}).}
\label{fig:sys_array_example}
\end{figure*}

This type of architectures 
are very suited for implementing these algorithms in low-power applications \cite{DBLP:journals/corr/abs-1906-11879}. Different architectures to implement this kind of algorithms on FPGAs were explored \cite{DBLP:journals/corr/abs-1711-05860,Wielgosz_2019,Lozito}. There is also a wide field of research focused exclusively on ways to improve the efficiency of the convolution operation in hardware \cite{10.1145/3020078.3021727,aleks2018hardwareefficient,10.1145/3020078.3021736}, since according to an analysis carried out by researchers from Tsinghua University, 98 \% of the mathematical operations performed by the nowadays most commonly used neural network architectures are done in the convolutional layers \cite{DBLP:journals/corr/abs-1712-08934}.
When this kind of digital designs are developed, certain characteristics are usually observed and taken into account as key aspects which need to be carefully analyzed, such as latency, partial additions storage, number of accesses to internal buffers, number of accesses to external buffers and data reuse, like spatial and/or temporal reuse \cite{10.1145/3020078.3021736}.

Another aspect to take into account is the structure of an optimal processing element. Works like \cite{10.1007/978-3-030-05677-3_16} \cite{isscc_2016_chen_eyeriss} have shown that this could be achieved by using a multiplier and an adder used as an accumulator. In order to maximize the internal data reuse, these works propose a matrix-wise interconnect for this processing elements (PEs) conforming a \textit{Systolic Array} \cite{1653825}, where the coefficients and the input \textit{pixels} are shared across adjacent elements. Finally, it is proposed that each PE should have a minimum internal control to be able to implement different kinds of operations and improve its reusability.

Most of current designs tend to use the algorithm known as \textit{im2col} to transform the convolution operation into a matrix multiplication, which can be easily accelerated using the increasingly bigger Systolic Arrays that are being developed nowadays. But this 
algorithm also adds a memory overhead because data needs to be rearranged in specific ways, which produce values duplication in memory. This is unsuitable for low resource devices where memory is a limited resource.

In contrast, this work presents a novel hardware accelerator architecture targeting resource constrained devices based on small and reusable Systolic Arrays. In order to validate its operation, the design was implemented on multiple FPGAs and a ResNet type neural network architecture was executed on it, achieving a maximum of 7.468 GOPs classifying images of the CIFAR-10 dataset \cite{cifar10}. Finally, the \textbf{Hardware Aware Pruning Method} (HAPM) is presented, a custom pruning technique which exploits the scheduling properties of this accelerator. When compared against the standard pruning technique, we demonstrate that networks trained with HAPM achieve a remarkable improvement of a 45 \% in the inference time per image without significant accuracy loss.
 
\section{Design of the hardware architecture}


\subsection{Optimization of the convolution operation}
\label{section:conv_opt}

\begin{algorithm}[!htbp]
	\scriptsize
	\caption{Pseudocode of a convolutional layer. Loops \ref{lst:x_axis}, \ref{lst:y_axis} and \ref{lst:z_axis} traverse the input matrix $i$ in their 3 dimensions ($N_{ix}$ and $N_{iy}$ already take into account the padding). Loops \ref{lst:group_selector}, \ref{lst:x_kernel} and \ref{lst:y_kernel} traverse the coefficient matrix $k$}\label{al:conv}
	\begin{algorithmic}[1]
		\State $o_{[..,..,..]} \leftarrow 0$\Comment{Output initialization}
		\For{$f\gets0$ to $N_{of}-1$ by $1$}\label{lst:group_selector}
			\For{$i\gets0$ to $N_{ix}-1$ by $s_{x}$}\label{lst:x_axis}
				\State $i_{o} \leftarrow i/s_{x}$
				\For{$j\gets0$ to $N_{iy}-1$ by $s_{y}$}\label{lst:y_axis}
					\State $j_{o} \leftarrow j/s_{y}$
					\For{$k\gets0$ to $N_{if}-1$ by $1$}\label{lst:z_axis}
						\For{$i_{k}\gets0$ to $N_{kx}-1$ by $1$}\label{lst:x_kernel}
							\For{$j_{k}\gets0$ to $N_{ky}-1$ by $1$}\label{lst:y_kernel}
								\State $o_{[i_{o},j_{o},f]} \leftarrow o_{[i_{o},j_{o},f]} + i_{[i+i_{k},j+j_{k},k]} \cdot k_{[i_{k},j_{k},k,f]}$ 
							\EndFor
						\EndFor
					\EndFor
					\State $o_{[i_{o},j_{o},f]} \leftarrow o_{[i_{o},j_{o},f]} + b_{[f]}$ \Comment{\textit{Bias} sum}
				\EndFor
			\EndFor
		\EndFor
	\end{algorithmic}
\end{algorithm}

The output of a convolutional layer can be calculated using a series of nested loops, as presented in algorithm \ref{al:conv}. But in a FPGA, these loops can be parallelized in different ways. In the work by Ma et. al. \cite{10.1145/3020078.3021736} the impact of 3 ways of implementing these cycles is analyzed, and this is how this strategies were used in this work:

\begin{itemize}
	\item \textit{Loop unrolling}: the loop at line \ref{lst:group_selector} was unrolled, and the design has the flexibility to unroll it fully or partially, depending on the amount of FPGA resources available. Loops at lines \ref{lst:y_axis},\ref{lst:x_kernel} and \ref{lst:y_kernel} were also partially unrolled.
	\item \textit{Loop tiling}: the design can be configured to select which layers have their coefficients stored in the internal buffers of the FPGA (Block RAM) and which layers have their coefficients stored in the external RAM. In the latter case, the coefficients are first brought to the internal buffers before starting to execute the layer.
\end{itemize}

\subsection{Core processing element}

As stated in algorithm \ref{al:conv}, the basic operation is a MACC operation (multiply and accumulate). Using works \cite{10.1007/978-3-030-05677-3_16,isscc_2016_chen_eyeriss} as a reference, a \textit{processing element} (PE) as seen in figure \ref{fig:sys_array_example_a} was implemented. This PE consists of a multiplier and an adder that can be used both to accumulate the output of itself with the multiplied value, or to accumulate the partial sum that comes from another cascaded PE.


\subsection{Computation units matrix}
\label{section:computation_units_matrix}

In order to reuse data while executing the convolution operation, a structure like the one presented in figure \ref{fig:sys_array_example_b} was proposed, composed of multiple cascaded PEs remembering a \textit{systolic array} of dimensions $CU_x$ and $CU_y$. By connecting the PEs in this way, it is possible to reuse the input data in multiple PEs: the green arrows indicate the reuse of the filter coefficients that convolve with the image, and the blue arrows indicate the reuse of the image data (or the input layer). This reuse decreases the number of memory accesses by implementing a \textit{spatial reuse} of data. It also eliminates the need to store special transformations of the input feature maps of a convolutional layer (as opposed to accelerators which use GEMM to calculate the output of a convolution, which incur in memory overheads because of this transformations): data is continuously streamed into the array in columns of $CU_h = CU_x + CU_y - 1$ values, traversing the input matrix from left to right. The partial results of the convolution are first accumulated into each PE and then forwarded to the top ones. By pipelining the arrival of the filter and data to each PE, this matrix is capable of computing the output of two $3\times3$ convolutions every 4 clock cycles.


\subsection{Matrix block}
\label{section:bloque_de_matrices}

This module is designed to instantiate $N_{CU}$ matrices in parallel as it is presented in figure \ref{fig:sys_array_example_c}. This is the module which actually executes the complete convolution operation. By selecting $N_{CU}$ one can control the desired parallelism level and the amount of resources used in the design. 


A particular design detail is highlighted regarding the buffers of each matrix. In contrast with the data, partial inputs and partial outputs buffers (FIFO buffers), the input coefficient buffer is a circular one, which allows the implementation of a \textit{temporal reuse}, saving substantial memory access time and energy. This is similar to the \textit{Weight Stationary} (WS) approach used by many Systolic Array based accelerators like Gemmini \cite{gemmini}.

In addition, this module can be synthesized with extra hardware elements that dynamically verify if the data loaded in the data or coefficient buffers is zero, in order to avoid losing clock cycles by performing unnecessary multiplications. This feature will be referred during the rest of the paper as \textit{Dynamic Sparsity Bypass} (\acronimo).

\begin{figure*}[!ht]
	\centering
	\includegraphics[width=0.8\textwidth]{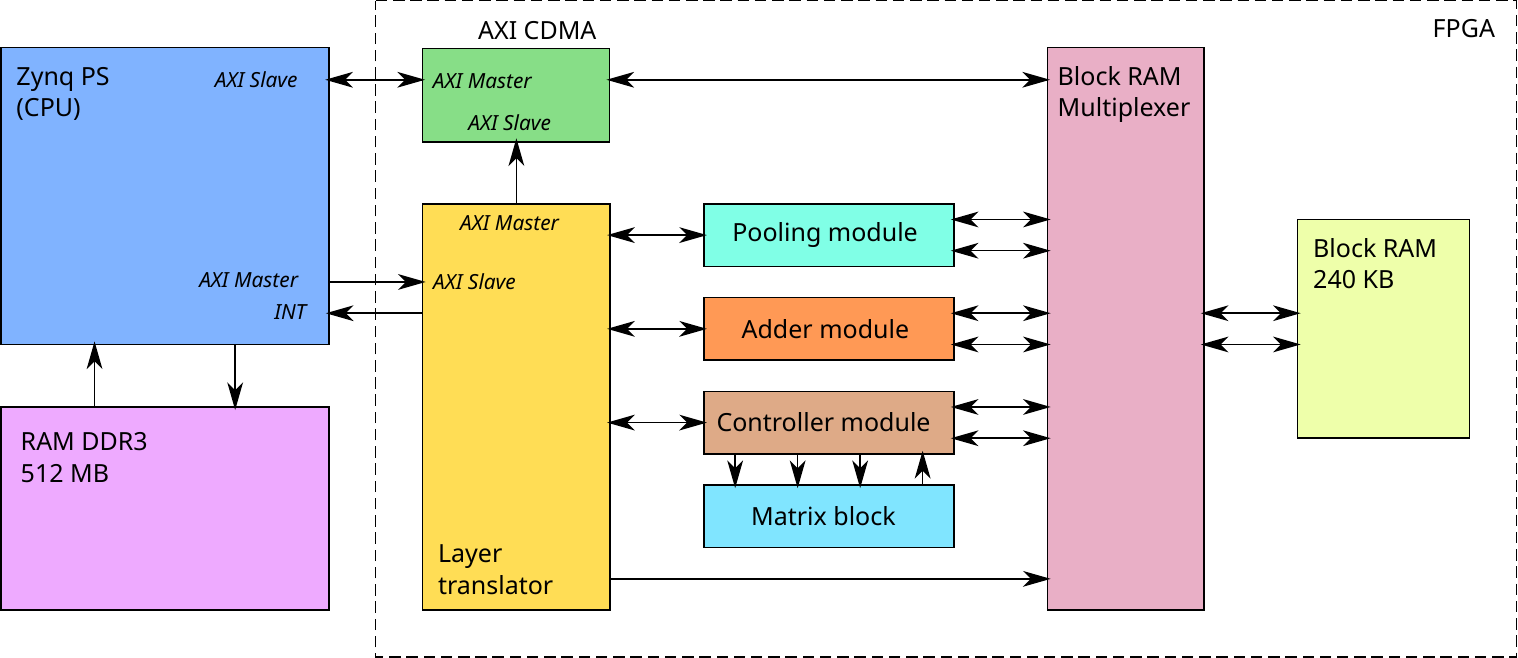}
	\caption{General block diagram of the proposed design}
	\label{fig:general}
\end{figure*}

\subsection{Convolution scheduling}
\label{section:conv_sch}

The order in which the data is dispatched to this last module, and the order in which the results of partial sums are read from it, is decided by the \textit{layer translator} module (see figure \ref{fig:general}). Every computation unit matrix receives the same input data but uses different convolution filters. Each one receives a patch of data to process, along with the appropriate filter and data to add to each output. This is called the \textit{scheduling} of the convolution (see algorithm \ref{al:conv_impl}). Lines \ref{lst:line:for_g}, \ref{lst:line:for_i} and \ref{lst:line:for_j} represent the loops over the input matrix, and line \ref{lst:line:for_dsp} represents the parallel processing of the matrix block module (section \ref{section:bloque_de_matrices}. Function \textbf{SysArray} represents the internal processing that is done in each computation unit matrix (section \ref{section:computation_units_matrix}).

\begin{algorithm}
	\scriptsize
	\caption{Pseudocode of the convolution schedule implemented on this work}\label{al:conv_impl}
	\begin{algorithmic}[1]
			\Function{ConvImpl}{$ i,k,b,s,N_{cu}$}
			
			\State $o \leftarrow 0$\Comment{Output matrix initialization}
			
			\State $t \leftarrow 0$\Comment{Temporal matrix initialization}
			
			\For{$f\gets0$ to $N_{of}-1$ by $N_{cu}$}\Comment{Group filter selector}
				\For{$g\gets0$ to $N_{if}-1$ by $1$} \label{lst:line:for_g}
					\For{$i\gets0$ to $N_{i}-1-N_{k}$ by $s$} \label{lst:line:for_i} 
						\State $p \leftarrow i/s$
						\State $r \leftarrow i+N_{k}-1$
						\For{$j\gets0$ to $N_{i}-1-4$ by $2$} \label{lst:line:for_j}
							\State $q \leftarrow j/s$
							\State $m \leftarrow j+3$ 
							\State $cols \leftarrow i[i:r,j:m,g]$
							\ParFor{$cu\gets0$ to $N_{cu}-1$ by $1$} \label{lst:line:for_dsp}
								\State $f_{cu} \leftarrow f+cu$
								\State $kernel \leftarrow k[:,:,g,f_{cu}]$
								\If{$f = 0$}
									\State $p_{1} \leftarrow b[f_{cu}]$
									\State $p_{2} \leftarrow b[f_{cu}]$
								\Else
									\State $p_{1} \leftarrow t[p,q]$
									\State $p_{2} \leftarrow t[p,q+1]$
								\EndIf
								\If{$f = N_{of}-1$}
									\State $o[p,q:q+1,f] \leftarrow$ \Call{SysArray}{$cols,kernel,p_{1},p_{2}$} \label{lst:line:output}
								\Else
									\State $t[p,q:q+1] \leftarrow$ \Call{SysArray}{$cols,kernel,p_{1},p_{2}$}
								\EndIf
							\EndParFor
						\EndFor
					\EndFor
				\EndFor
			\EndFor
			
			\State \Return output
			
			\EndFunction
			
			\Function{SysArray}{$i_{[N_{k},4]}$,$k_{[N_{k},N_{k}]}$,$presum_{1}$,$presum_{2}$}
				\State $out_{1} \leftarrow 0$
				\State $out_{2} \leftarrow 0$
				
				\For{$i\gets0$ to $N_{k}-1$ by $1$}
					\For{$j\gets0$ to $N_{k}-1$ by $1$}
						\State $out_{1} \leftarrow out_{1} + i_{[i,j]} \cdot k_{[i,j]}$
						\State $out_{2} \leftarrow out_{2} + i_{[i,j+1]} \cdot k_{[i,j]}$
					\EndFor
				\EndFor
				
				\State $out_{1} \leftarrow out_{1} + presum_{1}$
				\State $out_{2} \leftarrow out_{2} + presum_{2}$
				
				\State \Return $[out_{1},out_{2}]$
				
			\EndFunction
		\end{algorithmic}
\end{algorithm}

By analyzing the scheduling policy described in algorithm \ref{al:conv_impl} and the design of the computation units matrix, it is possible to determine the theoretical minimum number of clock cycles that this design may take to perform the complete calculation of a convolutional layer using equation (\ref{eq:ciclos_min}), taking into account: the parameters of the layer and the number and size of the computation units matrices instantiated in parallel. 
This equation assumes that: (a) each matrix always has data ready to be processed in its input FIFO buffers, and (b) at the same time the output buffers always has enough space to save the processed data.
\begin{IEEEeqnarray}{l}
		min_{cycles} = N_{valid}\cdot p_x\cdot p_y\cdot N_{if}\cdot ratio
		\label{eq:ciclos_min} \\
		p_x = (N_{i_x}-k_{o_x})/s_x \\
		p_y = ceil(G_{k_{y}}/G_{CU}) \\
		ratio = N_{of}/N_{cu} \\
		G_{cu} = floor((CU_{h}-k_{o_y})/s_y) \\
		G_{ky} = (N_{i_y}/k_{o_y}) - s_y \\
		k_{o_{x/y}} = max(abs(N_{kx/y}-s_{x/y}),1)
\end{IEEEeqnarray}
	
\begin{itemize}
	\item $N_{valid}$: the number of cycles that it takes for one matrix to have valid data available at its output. As it was mentioned in section \ref{section:computation_units_matrix}, this number is 4 for this version.
	\item $p_x$: the amount of groups of width equal to $N_{k}$ that each matrix needs to process.
	\item $p_y$: the amount of groups of height $CU_{h}$ that each matrix needs to process.
	\item $ratio$: the relation between the amount of output channels of the layer and the amount of instantiated computation matrices\footnote{This ratio should be a natural number}.
	\item $G_{cu}$: how many kernel windows can each matrix process at the same time.
	\item $G_{ky}$: how many kernel windows need to be processed.
	\item $k_{o_{x/y}}$: overlap of one kernel window with its neighbors on the x or y axis (if $N_{kx/y} = s_{x/y} \rightarrow k_{o_{x/y}} = 1$ for numerical stability of the other equations).	
\end{itemize}

As a numeric example, lets take a design with $N_{CU} = 12$, $CU_x = 2$ and $CU_y = 3$, which needs to calculate a convolutional layer with kernel sizes $k_x = k_y = 3$, strides $s_x = s_y = 1$ and $N_{of} = 12$ over an input matrix with sizes $N_{i_x} = N_{i_y} = 32$ and $N_{if} = 12$. By inserting this values in equation \ref{eq:ciclos_min}, this convolution should be processed in 12288 clock cycles.

\subsection{Block diagram of the entire architecture}
\label{ref:block_diagram}

Figure \ref{fig:general} shows all the modules interconnected with each other. A brief description of each of one them is presented:

\begin{itemize}
		\item \textbf{CPU/DDR3 DRAM:} this Zynq 7000 resources are used to run a Linux operating system. It's main function  is to obtain the image to be analyzed, notify the hardware through a driver that must start processing it, and wait for the result.
		\item \textbf{CDMA:} Performs data transfers between the external memory and the internal memory of the FPGA.
		\item \textbf{Computation units:} already described in section \ref{section:conv_opt}.
		\item \textbf{Principal modules:} the controller, adder and pooling modules each represent one layer of a classical convolutional neural network. They obtain the necessary data from the internal memory, perform the operation (internally or by sending the data to the matrix block module, section \ref{section:bloque_de_matrices}) and save the result to memory again.
		\item \textbf{Layer translator:} this module owns the knowledge about the architecture of the neural network to be executed (that is, the layers sequence, the types of each layer, the specific parameters of each one of them, etc.). It coordinates the operation of all the other modules. This allows to make the design configurable, selecting the network to run during the implementation step \footnote{Future versions should allow this to be dynamically configured}.
\end{itemize}

\begin{figure*}[ht]
	\centering
	\begin{subfigure}[b]{0.45\textwidth}
		\begin{tikzpicture}[]
			\begin{axis}[xlabel=Iteration,ylabel=Accuracy,y unit=\%,legend pos=south east, ymin=0, ymax=100, height=0.16\textheight, width=0.9\textwidth]
				\addplot [color=red] table[x=Step,y=Value,col sep=comma] {data/01/train-tag-epoch_accuracy.csv};
				\addplot [color=blue] table[x=Step,y=Value,col sep=comma] {data/01/validation-tag-epoch_accuracy.csv};
				\addlegendentry{Training}
				\addlegendentry{Validation}
			\end{axis}
		\end{tikzpicture}
		\caption{Model 1}
		\label{gra:primerversion_entrenamiento}
	\end{subfigure}
	\hfill
	\begin{subfigure}[b]{0.45\textwidth}
		\begin{tikzpicture}[]
			\begin{axis}[xlabel=Iteration,ylabel=Accuracy,y unit=\%,legend pos=south east, ymin=0, ymax=100, height=0.16\textheight, width=0.9\textwidth]
				\addplot [color=red] table[x=Step,y=Value,col sep=comma] {data/02/train-tag-epoch_accuracy.csv};
				\addplot [color=blue] table[x=Step,y=Value,col sep=comma] {data/02/validation-tag-epoch_accuracy.csv};
				\addlegendentry{Training}
				\addlegendentry{Validation}
			\end{axis}
		\end{tikzpicture}
		\caption{Model 2}
		\label{gra:cuantizacion_entrenamiento}
	\end{subfigure}
	\hfill
	\begin{subfigure}[b]{0.45\textwidth}
		\begin{tikzpicture}[]
			\begin{axis}[xlabel=Iteration,ylabel=Accuracy,y unit=\%,legend pos=south east, ymin=0, ymax=100, height=0.16\textheight, width=0.9\textwidth]
				\addplot [color=red] table[x=Step,y=Value,col sep=comma] {data/03/train-tag-epoch_accuracy.csv};
				\addplot [color=blue] table[x=Step,y=Value,col sep=comma] {data/03/validation-tag-epoch_accuracy.csv};
				\addlegendentry{Training}
				\addlegendentry{Validation}
			\end{axis}
		\end{tikzpicture}
		\caption{Model 3}
		\label{gra:pruning_entrenamiento}
	\end{subfigure}
	\hfill
	\begin{subfigure}[b]{0.45\textwidth}
		\begin{tikzpicture}[]
			\begin{axis}[xlabel=Iteration,ylabel=Accuracy,y unit=\%,legend pos=south east, ymin=0, ymax=100, height=0.16\textheight, width=0.9\textwidth]
				\addplot [color=red] table[x=Step,y=Value,col sep=comma] {data/04/train-tag-epoch_accuracy.csv};
				\addplot [color=blue] table[x=Step,y=Value,col sep=comma] {data/04/validation-tag-epoch_accuracy.csv};
				\addlegendentry{Training}
				\addlegendentry{Validation}
			\end{axis}
		\end{tikzpicture}
		\caption{Model 4}
		\label{gra:grouped_pruning_entrenamiento}
	\end{subfigure}
	\caption{Training and validation accuracy scores curves of the trained models}
	\label{gra:training_curves}
\end{figure*}
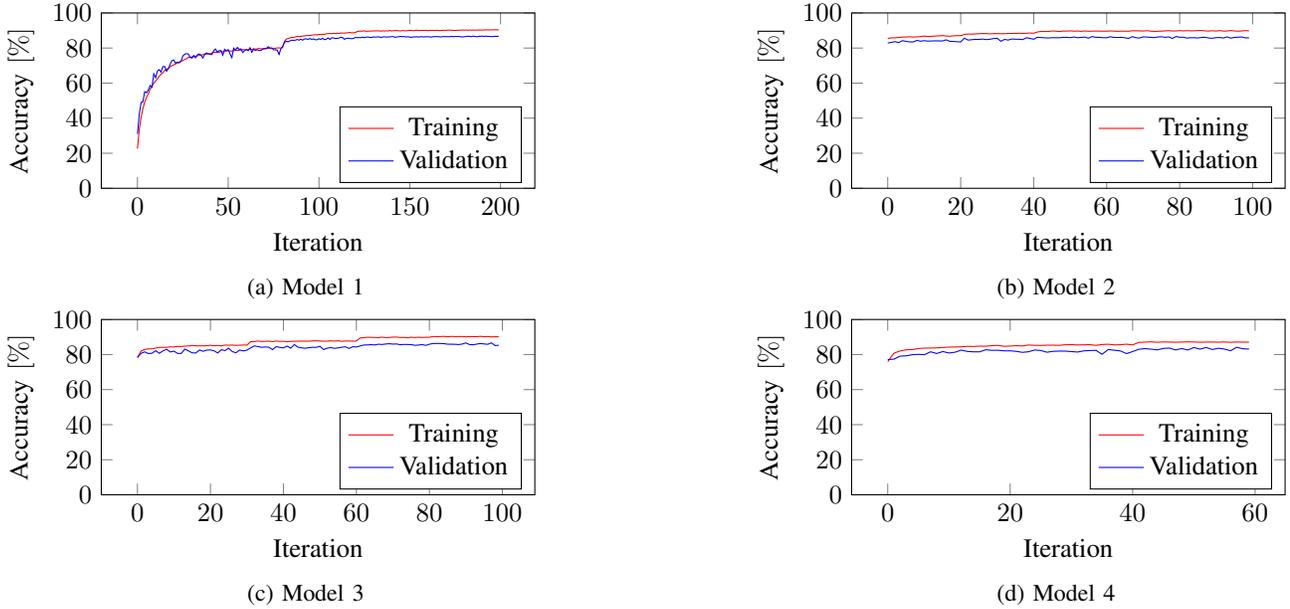

\section{Hardware Aware Pruning Method}

A common method when using convolutional neural networks in inference mode is to prune the weights of the network, in order to reduce its memory footprint. Several different approaches exist \cite{liang2021}: pruning can be done in an unstructured manner, per layer, per channel; the weights to prune can be chosen based on its absolute value or using the gradient to determine how much every weight contributes to the output; pruning can also be applied gradually (pruning fixed amounts of weights every step) or in a single step. 

The problem of all this approaches is that the remaining weights could be arranged in such a fashion that it would not be possible to speed that network using the already designed hardware accelerator. It is not enough to add sparsity to the weights of a network: this must be done in a structured way so that the hardware accelerator can take advantage of this sparsity, if it has the necessary logic to do so. This is why this paper presents our Hardware Aware Pruning Method (HAPM), whose pseudocode can be seen in algorithm \ref{al:hapm}.

\begin{algorithm}[H]
	\scriptsize
	\caption{Pseudocode of the HAPM}\label{al:hapm}
	\begin{algorithmic}[1]
		\State{Analyze schedule to form groups, based on which kernels are processed together}
		\State{$unpruned \leftarrow ...$} \Comment{Initialize with previous step groups}
		\State{$pruned \leftarrow empty$} 
		\State{$sparsity \leftarrow ...$} \Comment{Set desired sparsity}
		\State{$g \leftarrow sparsity*len(unpruned)/epochs$} \Comment{Groups to prune each epoch}
		\For{epoch in epochs} \Comment{Retraining}
		\State Sort groups in $unpruned$ by sum of abs values in asc. order
		\State Move the first $g$ groups from $unpruned$ to $pruned$
		\State Prune all groups in list $pruned$
		\State Continue basic epoch training loop
		\EndFor
	\end{algorithmic}
\end{algorithm}

First, the scheduling of the hardware accelerator should be analyzed, in order to detect which weights of a convolutional layer are processed in parallel. This are grouped together and each group is from now on treated as a single entity. After the user selects the desired level of group sparsity of the network, the retraining starts. At the beginning of each epoch, the unpruned groups are sorted into a list in ascending order according to the sum of the absolute values of the weights that conform them (we used this as a scoring method because it is widely accepted that the smaller the weights of a filter, the less this filter contributes to the output of the layer \cite{10.1145/3144789.3144803}). A certain amount of groups is selected from the start of this list and all the weights that are part of this groups are pruned. Then the basic training loop of the epoch continues. By repeating this steps before each epoch starts, we end up with a structured pruned network.

\section{Materials and methods}

\noindent In this work, a Zybo and a Zedboard boards were used to carry out the development. Each PE within the hardware design was implemented using one of the dedicated DSP48E1 found within all Xilinx 7-series FPGAs.

\subsection{Validation}

In order to validate the operation of the developed hardware, a neural network based on the ResNet architecture \cite{resnet1} composed of 21 convolutional layers was trained to classify images from the CIFAR-10 dataset \cite{cifar10}. This dataset contains 60,000 images of $32\times32\times3$ pixels, arranged in 10 mutually exclusive categories, which are separated into 50,000 images to train the network and 10,000 images to validate the performed training.

\begin{figure*}[ht]
	\centering
	\hspace*{-1.3cm}
	\begin{tikzpicture}
		\begin{axis}[
			ylabel=Zero weights percentage,y unit= \%,
			width=8in,
			height=2.0in,
			bar width=0.26cm,
			ybar,
			xtick=data,
			ymax=140,
			legend style={legend columns=-1},
			legend style={at={(0.24,0.95)}},
			symbolic x coords={
				0,
				1,
				2,
				3,
				4,
				5,
				6,
				7,
				8,
				9,
				10,
				11,
				12,
				13,
				14,
				15,
				16,
				17,
				18,
				19,
				20}]
			\addplot table[x=layer,y=sparsity,col sep=comma] {./data/grouppruned_model_sparsity_per_layer.csv};
			\addplot table[x=layer,y=sparsity,col sep=comma] {./data/pruned_model_sparsity_per_layer.csv};
			\addlegendentry{HAPM (4)}
			\addlegendentry{Pruning (3)}
		\end{axis}
	\end{tikzpicture}
	\caption{Sparsity per layer at the end of the training for models 3 and 4. Notice how our method chooses to almost suppress some layers, while keeping others practically intact.}
	\label{gra:grouppruned_sparsity_per_layer}
\end{figure*}
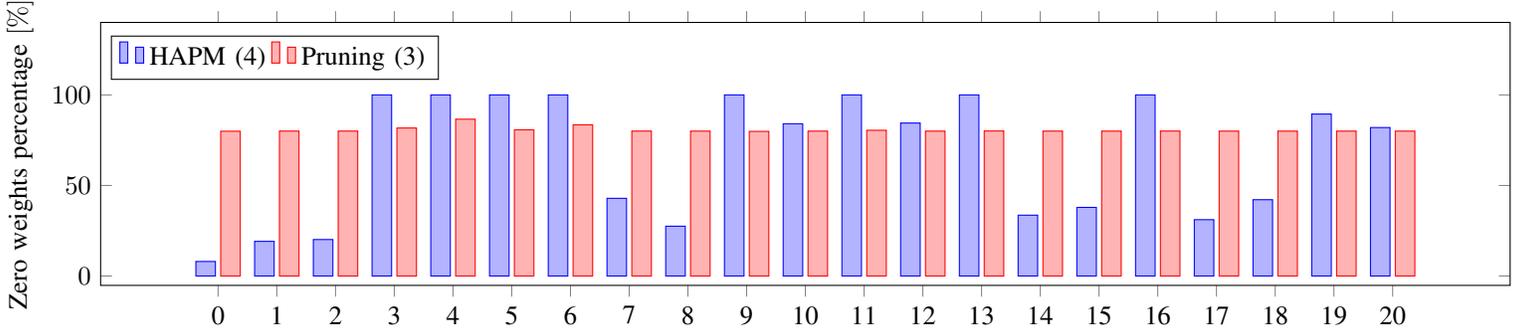

\subsubsection{Training}

In each trained version (see table \ref{tab:resultados_entrenamiento} and figure \ref{gra:training_curves}), the following techniques were used:
\begin{itemize}
 \item A variable learning factor was applied as the training progressed.
 \item The class \href{https://keras.io/api/callbacks/reduce_lr_on_plateau/}{ReduceLROnPlateau} of the Keras framework was used to also dynamically modify the learning factor.
 \item Since the data set is limited to 50,000 images, data augmentation techniques were used to expand the data set during training. For this, the class \href{https://keras.io/api/preprocessing/image/}{ImageDataGenerator} of the Keras framework was used. 
\end{itemize}

\begin{table}[H]
	\centering
	\scriptsize
	\begin{tabular}{cccccc} \toprule
		& Framework & Representation & Pruning & Accuracy [\%] \\
		\midrule
		1 & Keras \cite{Keras} & 32 bit floating point & - & 86,82 \\
		2 & QKeras \cite{QKeras} & 8 bit fixed point & - & 86,56 \\
		3 & QKeras \cite{QKeras} & 8 bit fixed point & Uniform \cite{zhu2017prune} & 86,65 \\
		4 & QKeras \cite{QKeras} & 8 bit fixed point & HAPM & 84,15 \\
		\bottomrule
	\end{tabular}
	\caption{PC training results}
	\label{tab:resultados_entrenamiento}
\end{table}
\begin{enumerate}
 \item This version was trained from scratch during 200 epochs (figure \ref{gra:primerversion_entrenamiento}) using the Keras framework \cite{Keras}, by representing all network weights as floating point values.
 \item This version was trained for 100 epochs using the weights pre-trained in version 1 (figure \ref{gra:cuantizacion_entrenamiento}), but quantizing them to a 8 bit fixed point representation using the QKeras framework \cite{QKeras}. The following fixed point arithmetic was chosen: Q2.5 for the network's coefficients  and Q3.4 for each layers output.
 \item This version was trained for 100 epochs using the weights pre-trained in version 2 (figure \ref{gra:pruning_entrenamiento}), by applying the \textit{uniform pruning} technique \cite{zhu2017prune}. A target pruning of 80\% was chosen for all layers (this means, 80\% of the weights of each layer were pruned).
 \item This version was also trained using the pre-trained weights of version 2, but it was trained for 60 epochs (figure \ref{gra:grouped_pruning_entrenamiento}), by applying the HAPM method. Figure \ref{gra:grouppruned_sparsity_per_layer} shows the sparsity per layer at the end of the training for this method vs the \textit{uniform pruning} trained in point 3. 
\end{enumerate}

{\footnotesize
	\begin{table*}[ht]
		\centering
		\footnotesize
		\begin{tabular}{ccccccccccccc}\toprule
			\multirow{3}{*}{Board} & \multirow{3}{*}{Frequency [MHz]} & \multirow{3}{*}{DSPs} & Model & \multicolumn{2}{c}{QKeras (2)} & & \multicolumn{2}{c}{Pruning (3)} & & \multicolumn{3}{c}{HAPM (4)} \\
			\cmidrule{5-6} \cmidrule{8-9} \cmidrule{11-13}
			& & & \acronimo?  & Yes & No & & Yes & No & & Yes & No & Yes\\
			& & & Depth data buffer & 8 & 8 & & 8 & 8 & & 8 & 8 & 32 \\
			\midrule
			\multirow{4}{*}{Zybo} & \multirow{5}{*}{70} & \multirow{5}{*}{72} & Accuracy [\%] & 81.88 & 81.88 & & 81.81 & 81.81 & & 79.72 & 79.72 &- \\
			& & & Total time [s] & 211.70 & 213.39 & & 211.43 & 213.39 & & 193.37 & 213.39 &- \\
			& & & Mean time per image [ms] & 21.17 & 21.34 & & 21.14 & 21.34 & & \textbf{19.34} & 21.34 &- \\
			& & & GOPs & 2.173 & 2.156 & & 2.176 & 2.156 & & \textbf{2.379} & 2.156 &- \\
			\midrule
			\multirow{4}{*}{Zedboard} & \multirow{5}{*}{100} & \multirow{5}{*}{72} & Accuracy [\%] & 81.49 & 81.88 & & 81.63 & 81.81 & & 79.42 & 79.72 & 79.72\\
			& & & Total time [s] & 124.93 & 126.33 & & 122.52 & 126.33 & & 70.04 & 126.32 & 64.54 \\
			& & & Mean time per image [ms] & 12.25 & 12.63 & & 12.25 & 12.63 & & 7.00 & 12.63 &\textbf{6,45} \\
			& & & GOPs & 3.682 & 3.641 & & 3.754 & 3.641 & & 6.568 & 3.641 &\textbf{7.127} \\
			\midrule
			\multirow{4}{*}{Zedboard} & \multirow{5}{*}{83.3} & \multirow{5}{*}{144} & Accuracy [\%] & 81.88 & 81.88 & & 81.81 & 81.81 & & 79.72 & 79.72 & 79.72 \\
			& & & Total time [s] & 107.10 & 107.81 & & 106.19 & 107.83 & & 66.70 & 107.81 & 61.60 \\
			& & & Mean time per image [ms] & 10.71 & 10.78 & & 10.62 & 10.78 & & 6.67 & 10.78 & \textbf{6.16} \\
			& & & GOPs & 4.295 & 4.227 & & 4.332 & 4.266 & & 6.897 & 4.267 & \textbf{7.468} \\
			\bottomrule
		\end{tabular}
		\caption{Results of different versions of the design executed on 2 different FPGA development boards.}
		\label{tab:resultados}
\end{table*}}

\subsection{Theoretical accelerator performance}

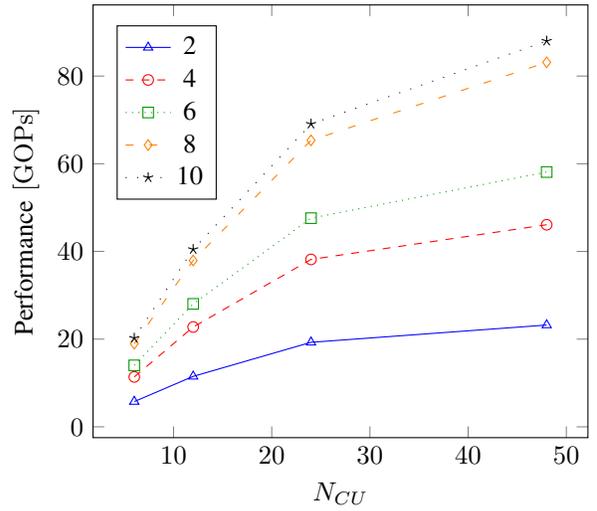
\begin{figure}[!ht]
	\begin{tikzpicture}[]
		\begin{axis}[
			scatter/classes={%
				2={mark=triangle},%
				4={mark=o},
				6={mark=square},
				8={mark=diamond},
				10={mark=star}}, 
			xlabel=$N_{CU}$,
			ylabel=Performance,
			y unit=GOPs,
			legend pos=south east, 
			height=0.3\textheight, 
			width=0.45\textwidth,
			legend image post style={mark indices={}},
			legend style={at={(0.25,0.55)},legend columns=1},
			legend entries={2,4,6,8,10}]
			\addlegendimage{blue,solid,mark=triangle,mark options={solid}}
			\addlegendimage{red,dashed,mark=o,mark options={solid}}
			\addlegendimage{green!60!black,dotted,mark=square,mark options={solid}}
			\addlegendimage{orange,loosely dashed,mark=diamond,mark options={solid}}
			\addlegendimage{black,loosely dotted,mark=star,mark options={solid}}
			\addplot [blue,scatter,solid,point meta=\thisrow{sysarraywidth},mark options={solid}] table[x=parallelsysarrays,y=gops,col sep=comma] {data/CNNAcceleratortheoreticalthroughput_width2.csv};
			\addplot [red,scatter,dashed,point meta=\thisrow{sysarraywidth},mark options={solid}] table[x=parallelsysarrays,y=gops,col sep=comma] {data/CNNAcceleratortheoreticalthroughput_width4.csv};
			\addplot [green!60!black,scatter,dotted,point meta=\thisrow{sysarraywidth},mark options={solid}] table[x=parallelsysarrays,y=gops,col sep=comma] {data/CNNAcceleratortheoreticalthroughput_width6.csv};
			\addplot [orange,scatter,loosely dashed,point meta=\thisrow{sysarraywidth},mark options={solid}] table[x=parallelsysarrays,y=gops,col sep=comma] {data/CNNAcceleratortheoreticalthroughput_width8.csv};
			\addplot [black,scatter,loosely dotted,point meta=\thisrow{sysarraywidth},mark options={solid}] table[x=parallelsysarrays,y=gops,col sep=comma] {data/CNNAcceleratortheoreticalthroughput_width10.csv};
		\end{axis}
	\end{tikzpicture}
	\caption{Theoretical performance of different configurations of the hardware accelerator for the chosen CNN, calculated at 100 MHz. Each plot represents one $CU_x$ parameter configuration.}
	\label{gra:theoretical_gops}
\end{figure}

By applying the equations presented in section \ref{section:conv_sch} to the neural network architecture selected in the previous section (which needs to do 0,046 GOP to completely process one image), the theoretical peak amount of GOPs that a certain design can achieve based on different parameterizations can be calculated. Figure \ref{gra:theoretical_gops} presents this results, by fixing $CU_y = 3$ and changing $CU_x$ and $N_{CU}$. The label of each point represent the amount of DSPs utilized by that particular design, assuming that each PE is composed of one DSP.

\subsection{Measurements}

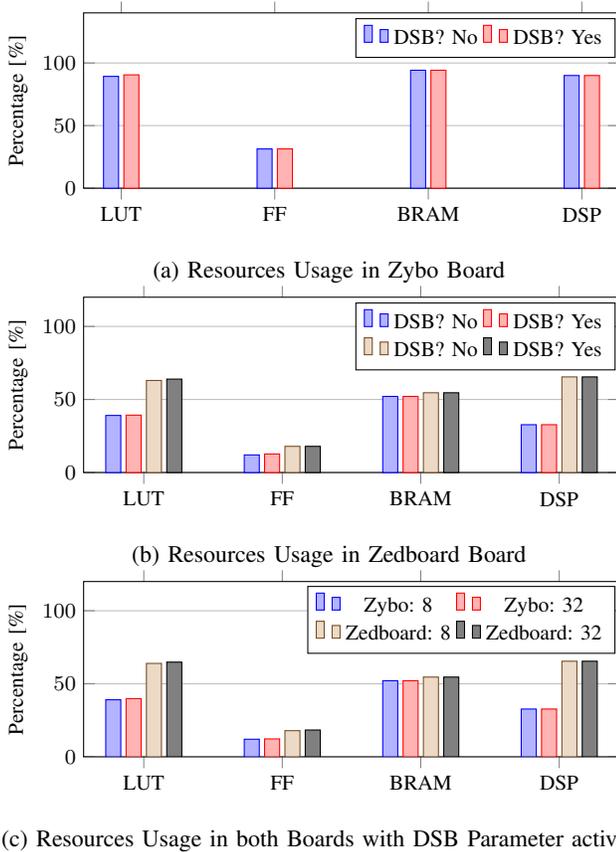
\begin{figure}[ht!]
\centering \footnotesize
 \begin{subfigure}[b]{0.48\textwidth}
  \begin{tikzpicture}[scale=1]
   \begin{axis}[ybar, bar width=.2cm, width=\textwidth, height=0.16\textheight,
   	legend style={at={(0.75,0.97)}, anchor=north, legend columns=-1}, symbolic x coords={LUT,FF,BRAM,DSP}, xtick=data, enlarge x limits={abs=0.5cm}, ylabel near ticks, ymin=0,ymax=140, ylabel={Percentage [\%]}, ymajorgrids=true,]
    \addplot table[x=resource, y=Zybo12-DSPSI]{\ResourceData};
    \addplot table[x=resource, y=Zybo12-DSPNO]{\ResourceData};
    \legend{\acronimo? No, \acronimo? Yes}
   \end{axis}
  \end{tikzpicture}
  \label{gra:Zybo_8}
  \caption{Resources Usage in Zybo Board}
 \end{subfigure}
 \begin{subfigure}[b]{0.48\textwidth}
  \begin{tikzpicture}[scale=1]
   \begin{axis}[ybar,bar width=.2cm,  width=\textwidth, height=0.16\textheight, legend style={at={(0.75,0.97)}, anchor=north, legend columns=2}, symbolic x coords={LUT,FF,BRAM,DSP}, xtick=data, enlarge x limits={abs=0.8cm}, ylabel near ticks, ymin=0,ymax=120, ylabel={Percentage [\%]}, ymajorgrids=true,]
    \addplot table[x=resource, y=Zedboard12-DSPSI]{\ResourceData};
    \addplot table[x=resource, y=Zedboard12-DSPNO]{\ResourceData};
    \addplot table[x=resource, y=Zedboard24-DSPSI]{\ResourceData};
    \addplot table[x=resource, y=Zedboard24-DSPNO]{\ResourceData};
    \legend{\acronimo? No, \acronimo? Yes,\acronimo? No, \acronimo? Yes}
   \end{axis}
  \end{tikzpicture}
  \label{gra:Zedboard_8}
  \caption{Resources Usage in Zedboard Board}
 \end{subfigure}
 \begin{subfigure}[b]{0.48\textwidth}
  \begin{tikzpicture}[scale=1]
   \begin{axis}[ybar, bar width=.2cm, width=\textwidth, height=0.16\textheight, legend style={at={(0.705,0.975)}, anchor=north, legend columns=2}, symbolic x coords={LUT,FF,BRAM,DSP}, xtick=data,enlarge x limits={abs=0.8cm}, ylabel near ticks, ymin=0,ymax=120, ylabel={Percentage [\%]}, ymajorgrids=true,]
    \addplot table[x=resource, y=Zedboard12-8]{\ResourceDataBis};
    \addplot table[x=resource, y=Zedboard12-32]{\ResourceDataBis};
    \addplot table[x=resource, y=Zedboard24-8]{\ResourceDataBis};
    \addplot table[x=resource, y=Zedboard24-32]{\ResourceDataBis};
    \legend{Zybo: 8,Zybo: 32,Zedboard: 8,Zedboard: 32}
   \end{axis}
  \end{tikzpicture}
  \label{gra:General_8}
  \caption{Resources Usage in both Boards with DSB Parameter activated}
 \end{subfigure} 
 \caption{Resource usage for the different implemented versions of the design. Figure a and b show the change in resource usage parameterized by the \acronimo \space parameter, for an input data buffer depth of 8 elements. Figure c shows the change in resource usage parameterized by the depth of the data buffer of each array of DSPs, when the \acronimo \space parameter is activated.}
 \label{fig:implementacion}
\end{figure}

For the execution of the trained networks on the designed hardware, multiple versions were implemented (for the resource usage reports of each version see figure \ref{fig:implementacion}), each one of them parameterized in a different way and with the internal RAM of the FPGA preloaded with the coefficients of one of the 3 quantized models presented in the previous section.

\pgfplotstableread[row sep=\\, col sep=&]{
	resource & Zybo12-DSPSI & Zybo12-DSPNO & Zedboard12-DSPNO & Zedboard12-DSPSI & Zedboard24-DSPNO & Zedboard24-DSPSI \\
	LUT & 89.32 & 90.52 & 39.28 & 39.07 & 63.95 & 63.01\\
	FF & 31.36 & 31.38 & 12.68 & 12.05 & 17.92 & 17.92\\
	BRAM & 94.2 & 94.2 & 52.1 & 52.1 & 54.6 & 54.6\\
	DSP & 90 & 90 & 32.73  & 32.73 & 65.45 & 65.45\\
}\ResourceData
\pgfplotstableread[row sep=\\, col sep=&]{
	resource & Zedboard12-8 & Zedboard12-32 & Zedboard24-8 & Zedboard24-32 \\
	LUT & 39.07 & 39.87 & 63.95 & 64.84\\
	FF & 12.05 & 12.27 & 17.92 & 18.37\\
	BRAM & 52.1 & 52.1 & 54.6 & 54.6\\
	DSP & 32.73  & 32.73 & 65.45 & 65.45\\
}\ResourceDataBis

\begin{figure}[!ht]
	\centering
	\begin{tikzpicture}
		\begin{axis}[
			ylabel=Performance,y unit= GOPs,
			width=0.45\textwidth,
			height=2.5in,
			bar width=0.26cm,
			legend style={at={(0.4,0.94)}},
			ybar,
			xtick=data,
			symbolic x coords={
				Zybo72DSPs,
				Zedboard72DSPs,
				Zedboard144DSPs}]
			\addplot table[x=model,y=gops,col sep=comma] {./data/results_2.csv};
			\addplot table[x=model,y=gops,col sep=comma] {./data/results_3.csv};
			\addplot table[x=model,y=gops,col sep=comma] {./data/results_4.csv};
			\addlegendentry{QKeras (2)}
			\addlegendentry{Pruning (3)}
			\addlegendentry{HAPM (4)}
		\end{axis}
	\end{tikzpicture}
	\caption{Improvement in the performance of the hardware accelerator for the selected CNN (higher is better). Notice the leap that can be achieved using the HAPM.}
	\label{gra:results}
\end{figure}
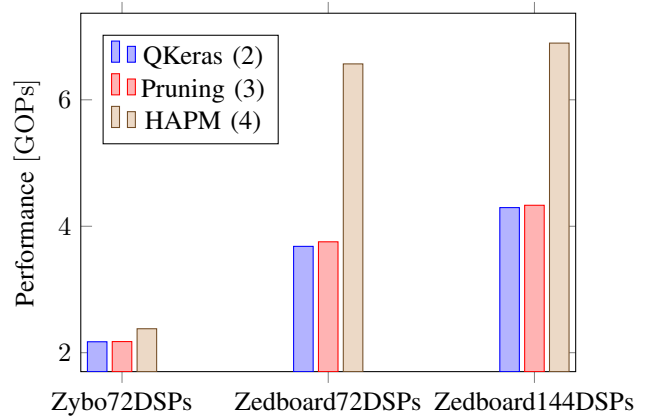

The 10,000 images of the CIFAR-10 test set were processed on the FPGA for the different implemented versions, obtaining the results presented in table \ref{tab:resultados} and figure \ref{gra:results}. The \textit{Total time [s]} row was measured from the point of view of the Linux kernel driver which communicates with the hardware accelerator: this time is the difference between the writing of the \textit{Start} AXI memory mapped register and the acknowledge of the \textit{finish} interrupt, both measured using the standard \textit{ktime\_get\_boot\_ns} Linux function. The \textit{Depth data buffer} row indicates the size of the FIFO buffers that communicate the \textit{Controller} module with the \textit{computation units}. 



\section{Discussion}
\noindent On one hand, for the \textit{QKeras} model (2), the activating of the \acronimo \space feature \textbf{does not add a significant improvement} to the mean inference time per image (this difference is as small as 0.79\%). This was expected and not at all surprising, since this model was not trained with any \textit{pruning} technique, and therefore the coefficients of its filters do not have any restriction applied to force them to zero and thus be able to take advantage of the \acronimo \space feature of the hardware.

It is also appreciated that the model trained with the uniform \textit{pruning} technique (model 3) adds a \textbf{slight improvement} when compared to the previous model, if versions with and without the \acronimo \space feature are compared (the difference is around a 3\% in the best case). However, this difference \textbf{is still very small}.

For the model trained with the HAPM (model 4), a \textbf{notable decrease} in the mean inference time per image of the design is observed, which is around \textbf{45 \%} in the best case, when feature \acronimo \space is activated. It is noteworthy that this model had been trained with a target pruning of only \textbf{50 \%}, and yet \textbf{obtained better results} than model 3 in terms of inference speed, at the cost of a slight loss of accuracy. This can also be seen in figure \ref{gra:results}, where the performance improvement across the different hardware implementations can be seen.

For model 4, the inference time per image was compared for a FIFO data buffer depth (for each computation unit) of 8 and 32 elements. If this buffer is too small, \textbf{idle states} appear in the internal state machines of the \textit{Controller} module, thus increasing the mean inference time per image. As shown in the table, a design implemented with 32-element deep data buffers achieves an 8 \% improvement in the inference speed. But on the other hand, this depth increase also brings with it an increase in the FPGA resource usage (figure \ref{fig:implementacion}). Therefore, although the improvement is \textbf{small} in this particular case, if there are \textbf{FPGA with more resources} available, a considerable improvement can be obtained by increasing the size of this buffer even more.

There is also a difference between the theoretical performance of the design, and the actual one once implemented. For example, the theoretical performance of the Zedboard model with 72 DSPs should be around 11 GOPs according to figure \ref{gra:theoretical_gops}, but in the best case, we achieved 7.468 GOPs. Part of this depends on the size of the FIFO buffers and the IDLE states, as described in the previous paragraph. But there are also scheduling issues, which should be addressed in following works. Because of the chosen scheduling policy, after the last channel of the input matrix of a convolutional layer is processed, the output data needs to be stored in its final memory position (see line \ref{lst:line:output} in algorithm \ref{al:conv_impl}). This data needs to be stored in a specific layout so that it can be used by the next layer without transformations. The problem is, that during this last storing of data, the output data of each computation units matrix needs to be stored in disjoint locations, and the entirety of the writing bus of the SRAM can not be used to pack together multiple writes. This generates back pressure on the output FIFOs and also delay the start of the processing of the next layer until all data is written to the SRAM.

\section{Conclusion}

\noindent We developed a CNN hardware accelerator small enough to be implemented for resource-constrained FPGAs. We demonstrated that HAPM can be used to accelerate the inference of a CNN on a custom hardware accelerator by almost $2\times$.

\section{Future work}

\noindent This work opens the doors to a large set of future research branches, where extensions and improvements to this hardware accelerator will be analyzed. On the one hand, the use of the DSP slices in SIMD mode will be investigated, and its impact in the throughput and performance of the design will be analyzed. On the other hand, the energy consumption analysis of these implementations and its relationship with the HAPM and the \acronimo \space feature are also pending topics worth investigating. In addition, it is of great interest to analyze with more detail why a network trained with the standard \textit{pruning} technique has so little improvement in the image processing time, and to review if the \acronimo \space characteristic can be improved for networks trained with this technique. On the other hand, in this version of the design, the coefficients of the neural network are stored and retrieved from the DRAM without any type of compression, which is not efficient when executing \textit{pruned} networks. The analysis of dynamic compression and decompression techniques for these data and their integration into this architecture are also of interest. Additionally, the scheduling and memory layout issues described in \textit{Discussion} should be addressed in order to leverage the difference between the theoretical performance and the real one. Finally, future work should demonstrate the flexibility of this design by running multiple different neural network architectures on more FPGAs.